\documentclass[12pt]{JHEP3} 

\usepackage[normalem]{ulem}
\usepackage{amsmath}
\usepackage{amsthm}
\usepackage{amssymb}
\usepackage{enumerate}
\usepackage{bm,longtable,enumerate}
\usepackage{amsmath,amssymb}

\newcommand{\beqa}{\begin{eqnarray}}
\newcommand{\eeqa}{\end{eqnarray}}

\newcommand{\beq}{\begin{equation}}
\newcommand{\eeq}{\end{equation}}

\title{BPS solution for eleven-dimensional
supergravity with a conical defect
configuration.}

\author{Cristine N. Ferreira$^{* \dagger} $  \\ 
$^* $N\'ucleo de Estudos em F\'{\i}sica, \\ Instituto Federal de Educa\c{c}\~{a}o,
Ci\^encia e Tecnologia Fluminense, \\
Rua Dr. Siqueira, 273, Campos dos
Goytacazes,\\ 
Rio de Janeiro, Brazil, CEP 28030-130 \\ \\
$^{\dagger} $    The Abdus Salam
International Centre for Theoretical Physics \\
Strada Costiera, 11
I - 34151 Trieste Italy  \\  \\
e-mail: crisnfer@pq.cnpq.br}

\abstract{In this work, we found the solution for the field equations of eleven-dimensional supergravity with a BPS conical topological defect configuration. We chose the solutions that corresponded to a M2-brane where the space-time presents the co-dimension two. The source of the topological defect is a sigma model where the brane tension is connected with the angular deficit. We analyzed the Killing spinor equations, a 3-form gauge potential and Einstein's equations, proving that it is possible to find a full solution for the system. We analyzed the near-horizon limit proving that it is possible to obtain an $AdS_4 \times S^1 \times \mathbb{E}^6 $   for certain values of the brane tension. We also discussed some applications in the strong coupling theory for condensed matter systems. }

\keywords{Conical Defect, Supergravity}

\preprint{}

\begin{document}

\section{Introduction}

The importance of a study on eleven-dimensional supergravity is that this theory presents the maximum dimension admitting supersymmetric extended objects \cite{Nahm:1977tg}. This theory  gives us a low-energy effective description of the M-theory. The full description of the M-theory remains unknown. Therefore, the eleven dimensional supergravity can be considered the main ingredient to study the super-unified theory. An important feature was the discovery of a network of dualities that can  map  the five superstring theories, I, IIA, IIB, heterotic $E_8 \times E_8 $ and SO(32), with the eleven-dimensional supergravity. 
For this reason it is commonly accepted that the M-theory is the mother of these five superstring theories. The other valid type of duality for the string/M-theory, involves the holographic principle which is crucial for investigating strongly coupled field theories 
where the most important  case is called  Anti-de-Sitter/Conformal Field Theory (AdS/CFT) correspondence \cite{Maldacena:1997re}. 
The AdS/CFT  opened a new branch of applications, for example  in $QCD$ \cite{Brower:2002er}, condensed matter systems \cite{Cassani:2011sv,Bu:2012zzb,Donos:2010tu,Hartnoll:2009ns,Bayona:2010sd}  and recently the $AdS_4$  black hole from M-theory \cite{Halmagyi:2013sla}. The  AdS/CFT conjecture establishes  the exact relation between  AdS supergravity  and Super Conformal Field Theories (SCFT). Maldacena \cite{Maldacena:1997re} showed that it is possible to establish this exact correspondence if we consider the large limit  of the color number,  $N \rightarrow \infty$, of certain SCFT in  the $'t Hooft$ coupling with  $\lambda \equiv g^{2}_{YM} N $ fixed and $\lambda >> 1$.
Considering the perturbative description of $N $ dynamical Dp-branes coupled to strings one can check that at  first order in the $\alpha\prime $ expansion, with all possible dimensionless parameter in game fixed,  decouples into a SCFT living in (p+1) dimension  and supergravity in flat space. By taking an analog limit $(g_s= g_{YM}^2)$ in the p-Brane solitonic description coming from the corresponding supergravity analysis, one argues that a pair of decoupled systems survive. The first being fluctuations in the near horizon limit of the given p-Brane solution and the second supergravity in flat space-time. In this way one is encouraged to conjecture the duality among the SCFT in (p+1) dimension coming from the string theory side and supergravity in the near horizon geometry of the p-Brane geometry, which is always $AdS_{p+2}\times S_{D-p}$ with D being the number of space time dimensions where the embedding string theory lives.
In the previous sketch some assumptions are implicitlly done  like for instance the validity of the supergravity and perturbative string descriptions of dynamical branes. But these are guarantied provided one works in the large $N $ with fixed $\lambda>>1$ limit previously mentioned before with the usual identification among both sides of the duality.
The field content of the theory is given by a metric field $g_{M N} $, the 3-form potential $A_{ M N O P} $ and their fermionic partners. In this theory    the AdS space time arises naturally as  $AdS_4\times S^7$ and $AdS_7 \times S^4$ \cite{Fund2}.  An important ingredient of the  string/M theory is the brane formulation. 
This is an effective formulation that encodes the low energy dynamics of the system of branes and forms. 
In this framework we can construct the M2-brane following the same prescription used for the Dp brane\cite{Duff:1994an}. A M2-brane has a three dimensional Poincar\'e invariant sector and a compact $SO(8)$ invariant group. The important feature of this solution is that it contains a near horizon limit $AdS_4 \times S^7$. By the use of a dual 4-form field strenght, we can also construct the M5-brane and their near horizon limit is $AdS_4\times S_7$.
Using this same logic, to understand  the M-theory,  we used a ABJM theory  formulated  by   Aharony, Bergman, Jafferis, and Maldacena \cite{Aharony:2008ug} that  are Super Conformal Chern-Simons (SCCS) gauge theories  with $ {\cal N} = 6 $ supersymmetry that are related to the M-theory  on $AdS_4 \times S^7/ \mathbb{Z}_k $ with N units of flux. This duality holds when we choose the gauge group $U(N)_k \times U(N)_{-k} $  with SCCS level $ k << N^{1/5} $.  
Otherwise,  when  $  N^{1/5} << k <<N  $,  the  most appropriated theory is the  dual description  in terms of the IIA string theory in $AdS_4 \times \mathbb{C}P^3 $ \cite{Chandrasekhar:2009ey}.  The later being also a very interesting regime of the duality with a lot of applications \cite{Giardino:2011dz,Kalousios:2009mp,Naghdi:2013vpa}.  Despite only having pointed out some  implications of the low energy limit of  the M-theory, there is no doubt of the importance of their study.
For these and others motivations, this work was dedicated to the development of a new class of solutions for eleven dimensional supergravity with a conical defect. This description required that the  space time contains a sector that presents  the  co-dimension two.  
Topological defects are expected to be formed during phase transitions in the early universe. 
The most studied,  because  of their stability, was the cosmic string. This defect is analog to a superconductor in condensed matter, for this reason, they are some times called flux tubes. These tubes appear in materials and when coupled with the gravitation are called cosmic strings.  
In the universe these structures were not observed, yet. Some effort has been devoted to their detection \cite{Movahed:2010zq,Movahed:2012zt}. These defects were extensively   studied in connection with  structure formation 
but now  their interest has been reborn with the superstring context.
Despite the great interest in the study of topological defects, such as cosmic strings in the superstring theory to be the interface within cosmology \cite{Copeland:2009ga} they are also important for the study of strongly coupled systems in condensed matter \cite{Vozmediano:2010zz,Cortijo:2006xs,Cortijo:2006mh} with a lot of contexts. 
 In analogy with the D7-branes, which contains this type of defect \cite{Bergshoeff:2007zz}, the formalism should be a M8 brane, however these types of branes are massive, and are not the aim of this work, for this reason we choose to work with  the split (3,2,6). We showed that it is possible to find a  BPS-like structure, with this type of construction.

The ${\cal N} =1$ supergravity model for eleven dimension space-time involves a set of massless fields  which carry a representation of supersymmetry. Since supersymmetry assigns to each bosonic degree of freedom a corresponding fermionic one,  we can obtain a relation between bosons and fermions  by supersymmetric transformations.  The action of the model,  invariant by the SUSY transformation,  contains only the metric field $g_{M N} $ and the three form potential $A_{ M N O} $ \cite{Cremmer:1978km}. The bosonic part of the low-energy action is given by
\begin{eqnarray}
S &=& {1 \over 4}\int d^D x  \sqrt{-g} R - {1 \over 48} \int d^Dx  \sqrt{-g} F_{ M N O P } F^{ M N O P  }  \nonumber \\
& & +{1 \over (12)^4} \int d^Dx \epsilon^{ MNOP QRST UVW} F_{ MNOP  } F_{ QRST } A_{UVW} \label{action1}
\end{eqnarray}
 where the capital letters index all space from $M = 0...10$.
This action is invariant in ${\cal N} =1$ supersymmetric  transformations given by the following equations
\begin{eqnarray}
\delta e_{\, \, \, M}^{ \hat N} &=& - i \bar \epsilon \Gamma ^{\hat  N} \psi_{M}  \\
\delta \psi_{M} &=& D_{ M } \epsilon - {1 \over 288} ( \Gamma ^{ \, \, \, \, \,  L O P Q }_{ M } + 8 \Gamma^{ O P Q }  \delta^{L  } _{M} ) F_{  L O P Q }\,  \epsilon = \bar D_{M} \epsilon \label{rarita} \\
\delta A_{M N O} &=& {3 \over 2 } \bar \epsilon  \gamma_{[ M N } \psi_{ O ]}.
\end{eqnarray}
We organized this work as follows. Section 2 is dedicated to  the construction of the Killing spinor equations as well as the formulation of BPS conditions. In Section 3, we discuss the brane action, which is a sigma model. We also analyze  the complete solution using the field equations for the 3-form gauge, we discuss  the equations of Einstein and  the topological conical configuration of the defect. The analysis of the  near horizon limit AdS   were presented in Section 4. Finally,  in Section 5, we present our main results with a discussion about a dimensional reduction of eleven dimensions to ten. We also  make  some analyses  about  the applications for  future works.

\section{ Killing spinors for the M2-brane with a (3,\underline{2},6) split. }
In this section,  we consider the  construction of the M2-brane BPS solution with a  (3,\underline{2},6) split.  The disposition of this brane,  according to the pattern, is\begin{equation}
\begin{array}{ccccccccccc}
M2: &  1 & 2 & \odot & \odot &\otimes &\otimes&\otimes&\otimes&\otimes &\otimes
\end{array} 
\end{equation}
The fields here only depend on the sector of the  co-dimension two  that was represented  by $\odot$.  The other transverse sector is represented by $\otimes $ that presents six coordinates.  We labeled this pattern as $(3, \underline{2},6)$, where the underlined number refers to the dependence  of the  fields.  In this split, the D=11 coordinates can be written as
\begin{equation}
X^M = (X^{\mu}, Y^m,Z^{\tilde m} ) \label{split1}
\end{equation}
where we labeled  $\mu =0,1$ and $ 2 $ as the Minkowiski space-time, $m = 1, 2 $ is the conical defect transverse coordinates and $ \tilde m = 1,...,6$ refers to the other transverse coordinates.  The metric is as follows 
\begin{equation}
ds^2 = e^{2 A(y_1,y_2)}(- dt^2 + dx_1^2 + dx_2^2) + e^{2 B(y_1,y_2)}(dy_1^2 +dy_2^2) + e^{2C(y_1,y_2)}\sum_{n=1}^{6} dz_ndz_n \label{metric1}
\end{equation}
we considered the ansatz of the three-form gauge field  as follows
\begin{equation}
 A_{\mu \nu \rho }(y_1,y_2) = \pm \, {c \over ^3 g} \epsilon_{\mu  \nu  \rho } \, e^{E(y_1,y_2)} \label{potential1}
\end{equation}
where $c $ is a constant that in this moment is considered  arbitrary and $ ^3g $ is the metric determinant   that gives us Levi Civita's tensor definition that in a gravitational context is $  \epsilon_{ \mu  \nu  \rho } \equiv g_{\mu \alpha}     g_{\nu  \beta}  g_{\rho  \lambda} \epsilon^{\alpha \beta  \lambda}      $. We considered  all other components such as  $A_{MNO} $,  the graviton and the gravitino $ \psi_M$ as  zero.  We can see that  $A$, $B$, $ C$ and $ E$ depend only on $y^m$. We chose this theory due to the fact that it is the simplest case where there is a conical defect in 11 dimensions.
In this split  there are four arbitrary functions ($A$, $B$, $C$ and $E$)  which are reduced to one due to the requirement of the field configuration, (\ref{metric1}) and (\ref{potential1}), preserve some unbroken supersymmetry. For this reason, there are  Killing spinors that satisfy the following equation\begin{equation}
\bar D_{M} \epsilon =0 \label{covariant1}
\end{equation}
where $ \bar D_{M} $ is the super covariant derivative appearing in gravitino's supersymmetric transformation.  We can write this (\ref{rarita}) in the following manner
\begin{equation}
\delta \psi_{M} = (D_{M} +      A_{M}^{(1)}  + A_{M}^{(2)}    ) \epsilon = \bar D_{M}  \epsilon \label{covariant2}
\end{equation}
The covariant derivative part involving the spin connection is given by $
D_{ M} \epsilon  =  \partial_{M} \epsilon - {1 \over 4} \omega_{M}^{ \, \, \, \hat A \hat B}\Gamma_{\hat A \hat B}  $,  where $M$ is the Lorentz index and $\hat A$ is the flat index. The convention is $e^{\, \, \, M}_{\hat A} e_{\hat B   \, M} = \eta_{\hat A \hat B}$ and $ e_{M}^{\hat A} e_{ N  \hat A} = g_{MN}$ .
The covariant derivative involving  the flux field contribution is as  follows
\begin{eqnarray}
\omega_{M \hat  A}^{\,\, \, \, \,  \,\, \, \, \,  \hat B} &=& e_{N}^{\,  \,\, \hat  B}e_{ \hat A}^{\, \, \,  O} \,\Omega^{N}_{\, \, \, \,   M  O}  - e_{ \hat A}^{\, \, \, O} \partial_{ M } e_{O }^{ \, \, \, \hat B}  \label{spinconexion2}\\
A_M^{(1)} & =&  -{1 \over 288} \Gamma ^{ \, \, \, \, \,  N O P Q }_{ M } F_{  N O P Q }\\
A_M^{(2)} &=&  {1 \over 36} \Gamma^{ O P Q }  \delta^{L  } _{M}  F_{  L O P Q }
\end{eqnarray}
 where $ F_{M N O P} =  4 \partial_{[M}A_{N O P]}$. The first term in (\ref{spinconexion2}) is the Christoffel symbol  that we labeled as $\Omega_{MNO}$ to differ from the Dirac Matrix where we originally use $ \Gamma$.
The Dirac Matrices,  $\Gamma_{ A }$, in $D=11 $ satisfy $ [\Gamma_{A}, \Gamma_{B}]_+ =2 \eta_{  A   B} $ and  the metric signature is $\eta_{ A  B} = diag \, (-, +,...+ ) $.
We consider the decomposition of the $\Gamma $ matrix, that respecting  the (3,2,6)  split, is shown as follows
\begin{equation}
\Gamma_A = ( \gamma_{\alpha} \otimes \Sigma_3 \otimes \Gamma_7\, , \mathbb{I} \otimes \Sigma_a \otimes \mathbb{I}\, , \mathbb{I} \otimes  \Sigma_3 \otimes \Theta_{\tilde a}),\label{gammasplit}
\end{equation}    
where $\gamma_{\alpha} $ ,  $\Sigma_a$ and $\Theta_{\tilde a}$ Dirac matrices correspond to $D=3$, $D=2$ and $D=6$  dimensions respectively, with 
\begin{eqnarray}
\Sigma_3 &\equiv &\Sigma_1 \Sigma_2 \label{casimir1}\\
\Gamma_7& \equiv& \Gamma_1 \Gamma_2 .... \Gamma_6. \label{casimir2}
\end{eqnarray}
%%%%
The  consistent form to write the spinor  is as follows  
\begin{equation}
\epsilon = \epsilon_1 \otimes \epsilon_2 (y_1,y_2) \otimes \epsilon_3 
\end{equation}
where $ \epsilon_1$ is a constant spinor of SO(1,2), $ \epsilon_2$ is a  spinor of co-dimension 2 sector and $\epsilon_3$ is a constant spinor of D=6 transverse sector. 
The super covariant derivatives with a three split  are as follows
\begin{equation}
\bar D_{\mu} = \partial_{\mu}  -{1 \over 2} \gamma_{\mu} e^{-A} \Sigma^a \partial_a e^A \, \Sigma_3 \Gamma_7 \mp {1 \over 6} \gamma_{\mu} e^{-3 A} \Sigma^m \partial_m e^E \label{killing1}
\end{equation}
\begin{eqnarray}
\bar D_{a} &=& \partial_{a}  +{1 \over 4}  e^{-B} (\Sigma_a \Sigma^m - \Sigma^m \Sigma_a)         \partial_m e^B \nonumber \\ 
& & \mp {1 \over 24} e^{-3 A}   (\Sigma_a \Sigma^m - \Sigma^m \Sigma_a)    \partial_m e^E \Sigma_3 \Gamma_7\nonumber  \\
& & \mp {1 \over 6} e^{-3A} \partial_a e^E \, \Sigma_3  \Gamma_7 \label{killing2}
\end{eqnarray}
\begin{eqnarray}
\bar D_{\tilde m} &=& \partial_{\tilde m} \, + \, {1 \over 4}  e^{- C}( \Theta_{\tilde m} \Sigma^a - \Sigma^a  \Theta_{\tilde m})\,  \partial_a e^{C} \nonumber \\
& & \mp {1 \over 24} e^{-3 A} ( \Theta_{\tilde m} \Sigma^a - \Sigma^a  \Theta_{\tilde m})\partial_a e^E \Sigma_3 \Gamma_7 \label{killing3}
\end{eqnarray}
The  constraint to preserve unbroken supersymmetry  is given by (\ref{covariant1})  considering (\ref{killing1}), (\ref{killing2}) and (\ref{killing3}) in (\ref{gammasplit}) it  leads to, with c=1,  a constraint for the metric  function A,
\begin{eqnarray}
- {1 \over 6} \gamma_{\mu} \Sigma^a \partial_a E \,  (1 \pm       \, \Sigma_3 \Gamma_7)  \epsilon =0 & \rightarrow & A= {1 \over 3} E . 
\end{eqnarray}
The constraint for the metric function B is given by 
\begin{eqnarray}
-{1 \over 24}   (\Sigma_a \Sigma^m - \Sigma^m \Sigma_a)         \partial_m E  \, (1 \pm       \, \Sigma_3 \Gamma_7)  =0 & \rightarrow & B = - {1 \over 6} E \\
 \partial_{m}  \mp {1 \over 6}  \partial_m E \, \Sigma_3 \Gamma_7 =0 & \rightarrow & \epsilon  = e^{-E/6} \epsilon_0, 
\end{eqnarray}
and finally the constraint for the function  C is
\begin{eqnarray}
-{1 \over 24}   (\Theta_{\tilde m} \Sigma^a - \Sigma^a \Theta_{\tilde m})         \partial_a E  \, (1 \pm       \, \Sigma_3 \Gamma_7)  =0 & \rightarrow & C = - {1 \over 6} E. 
\end{eqnarray}
then, we can write the metric in the form
\begin{equation}
ds^2 = e^{{2 \over 3} E} \Big[ - dt^2 + dx_1^2 + dx_2^2 + e^{-E}(dy_1^2+ dy_2^2) \Big] + e^{- {1 \over 3}E} \sum_{\tilde m =1}^{6} \, dz_{\tilde m} dz_{\tilde m} \label{metric2}.
\end{equation}
In this calculation we omit the tensorial product and the  $\pm$ signs are correlated with our gauge potential ansatz (\ref{potential1}).
The complete solution to the spinor (\ref{split1}) is given by
\begin{equation}
\epsilon = e^{-E/6}  \epsilon_0 \label{spinorsolution1}
\end{equation}
\begin{equation}
\mathbb{I} \otimes \Sigma_{3} \otimes \Gamma_{7}  \, \epsilon_0 =  \mp \,  \epsilon_0  =   \left\{\begin{array}{cc} 
\Sigma_3 \epsilon_{0 2} =  \mp \epsilon_{02}& \\
\Gamma_7 \epsilon_{03}=  \mp \epsilon_{03}&
\end{array}\right. \label{branesolution1}
\end{equation}
where $\epsilon_{02}$ is a constant co-dimension two spinor and $ \epsilon_{03}$ is a co-dimension six spinor.
We showed in this section that our propose satisfies these Killing spinor equations. These equations are responsible for the BPS bound that  guaranties the stability of our solution.  In the next  Section,  we introduced the brane action. We studied the solution for the function E by analyzing the gauge field and Einstein equations. 

\section{The brane's configuration analysis }
In the last section we showed that it is possible to find a stable co-dimension two solution analyzing the Killing Spinor equations. In this section we discussed the source of the fields of our system considering the brane action.  
The brane action is compatible with eleven dimensional supergravity  \cite{Bergshoeff:1987dh} and can be written as
 \begin{eqnarray}
 S_M &=& \mu \int d^3 \xi \Big(- {1 \over 2} \sqrt{- \gamma} \gamma^{i j} \partial_i X^M \partial_j X^N g_{MN} + {1 \over 2} \sqrt{- \gamma} \nonumber \\
 & & \pm {1 \over 3 !} \epsilon^{i j k} \partial_i X^M \partial_j X^N \partial_k X^O A_{M N O} \Big)\label{brane1}
 \end{eqnarray}
where $\mu $ is the tension of the supermembrane. This action is invariant to ${\cal N}=1 $ supersymmetry transformations given by the following equations
 \begin{eqnarray}
 \delta \Psi_M &=& \bar D_M \epsilon(X) \label{covariant3}\\
 \delta \theta &=& (1 \pm \Gamma) \kappa(\xi) + \epsilon(X) 
 \end{eqnarray}
where $\Psi_M$ is the gravitino and $\theta$ is the fermionic coordinate, $\kappa(\xi)$ is the Siegel symmetry parameter and the covariant derivative $\bar D_M$  (\ref{covariant3}) is the same as (\ref{covariant2}) with a $\Gamma$ given the following
 \begin{equation}
 \Gamma \equiv {1 \over 3! \sqrt{-\gamma} } \epsilon^{ijk} \partial_i X^M \partial_j X^N \partial_k X^P \Gamma_{MNP}
 \end{equation}
with the BPS bound constraints as
\begin{equation}
\bar D_M \epsilon =0 \,\,\,\,\, \Gamma \epsilon = \mp \epsilon
\end{equation} 
 The  antisymmetric tensor field equation is given by the variation of the action (\ref{action1}) in relation with the potential $ A_{M N O}$ and results in
\begin{eqnarray}
\partial_M(\sqrt{-g} F^{M N O P})  + {1 \over 1152} \epsilon^{ N O P M_1 ... M_{8}} F_{M_1... M_4} F_{M_5 ... M_{8}} \nonumber \\
= \mp \,  \kappa \mu \int d^3 \xi \, \epsilon^{i j k }  \partial_i X^ N \partial_j X^O \partial_k X^P \delta^{11}(x-X)\label{gaugequation1}.
\end{eqnarray}
The equation of the membrane  field  is
\begin{eqnarray}
\partial_i(\sqrt{-\gamma} \gamma^{ij} \partial_j X^N g_{MN}) + {1 \over 2} \sqrt{- \gamma} \,\gamma^{ij} \, \partial_i X^N \partial_j X^P \partial_M g_{NP} \nonumber \\
\pm \, {1 \over 3!} \,  \epsilon^{i j k} \partial_i X^N \partial_j X^O \partial_k X^P F_{M N O P} =0
\end{eqnarray}
where $ \gamma_{ij} = \partial_i X^M \partial_j X^ N g_{MN}    $.  We considered the static gauge choice as 
\begin{equation}
X^{\mu} = \xi^{\mu}, \, \, \, \, \, \, \mu= 0,1 2, \label{branesplit1}
\end{equation}
and the transverse sector as
\begin{eqnarray}
Y^m &=& constant , \label{branesplit2} \\
Z^{\tilde m} &=& constant  \label{branesplit3}
\end{eqnarray}
It is easy to see that  (\ref{branesplit1}), (\ref{branesplit2}) and (\ref{branesplit3}) are compatible with our solution, where $\Gamma_{MNO}$ is reduced to $ \Gamma_{\mu \nu \gamma} $ by using  the duality relation, $ \Gamma_{\mu \nu \kappa} = \epsilon_{\mu \nu   \kappa a b \tilde a_1... \tilde a_6} \Sigma^a \Sigma^b \Gamma^{\tilde a_1} ...\Gamma^{\tilde a_6}$,  that with the split  (\ref{gammasplit})  giving us
\begin{equation}
\Gamma = \mathbb{I} \otimes \Sigma_3 \otimes \Gamma_7
\end{equation}
where $\Sigma_3$ and $\Gamma_7$ are defined in the eqs. (\ref{casimir1}) and (\ref{casimir2}). 
We can  compute the solution of the gauge field equation (\ref{gaugequation1}) considering the brane action (\ref{brane1}),  the metric (\ref{metric2}) and the action (\ref{action1}). These, together with the ansatz (\ref{potential1}) give us the only contribution for the equation of motion  as
\begin{eqnarray}
  \delta^{mn}  \partial_{m}  \partial_n \, e^{-E(y_1,y_2)}  = -16\,   G \mu \, \delta^2(y_1,y_2) \label{gaugeequation1}
\end{eqnarray}
In conclusion we obtained the solution as the following
\begin{equation}
e^{-E(r)} = 1- 8 G \mu \ln({r\over r_0}) \end{equation}
where $r_0$ is the minimum  radius of circle and  $ r = \sqrt{y_1^2 + y_2^2}$.

Now, we will analyze the Einstein equation from  the  conical defect  in eleven dimensional supergravity. We considered  the fermion field as zero and  we worked on the bosonic part  of the $D= 11$ theory.  The Einstein equation can be written as\begin{equation}
R_{MN} - {1 \over 2} g_{MN} R   -   {1 \over 12} ( F^M_{\, \, \, OPQ} F^{N O P Q} - {1 \over 8} g^{MN} F_{OPQR}F^{OPQR} ) = 8 \pi G T_{MN},
\end{equation}
where the energy momentum tensor $T_{MN}$ is given by
\begin{eqnarray}
T^{MN} &=& -  \mu   \int d^3 \xi \epsilon^{i j k} \partial_i X^M \partial_j X^N {\delta^{11}(x-X) \over \sqrt{- g}}
\end{eqnarray}
\begin{equation}
\tilde G_{M N}=G_{MN} -{1 \over 12} [F_{M O P Q  }F_{N}^{\, \, \,  O P Q} -{1 \over 8} g_{M N } F_{O P Q L }  F^{O P Q L }] = 8 \pi G \, T_{MN}, 
\end{equation}
we redefined the Einstein tensor as $  \tilde G_{M N} $ only for commodity where we have identified three different parts. One of these  is given by the Minkowiski sector,
\begin{eqnarray}
G_{\mu \nu} = -{1 \over 4} \delta^{ab} \Big[ \partial_a E(y_1,y_2)\partial_bE(y_1,y_2) - 2\partial_a \partial_b E(x_1,x_2) \Big]   e^{E} \,  \eta_{\mu \nu }
\end{eqnarray}
the other part corresponds to the co-dimension two sector,
\begin{eqnarray}
G_{ab} ={1 \over 4} \left( \begin{array}{cc}  (\partial_{a} E(y_1,y_2))^2 -  (\partial_{b} E(x_1,x_2)   )^2 &  2 \partial_{a} E(y_1,y_2)  \partial_{b} E(y_1,y_2)  \\
2 \partial_{a} E(y_1,y_2)  \partial_{b} E(y_1,y_2) & (\partial_{b} E(x_1,x_2)   )^2 - (\partial_{a} E(y_1,y_2))^2
\end{array} \right)
\end{eqnarray}
and the last one is given by the co-dimension six sector,
\begin{equation}
G_{\tilde a \tilde b }= -{1 \over 4} \delta^{ab}  \partial_a E(y_1,y_2)\partial_bE(y_1,y_2) \, \eta_{\tilde a \tilde b}
\end{equation}
Using the definition of flux we can write the Einstein equation for the defect where the non zero components for the $\tilde G_{MN}  $ are
\begin{eqnarray}
\tilde G_{\mu \nu} &=& -{1 \over 2}e^{2 E(y_1,y_2)} \delta^{mn}  \partial_{m}  \partial_n \,e^{ -E(y_1,y_2)}  \eta_{\mu \nu}   \nonumber \\
& = & 8\,  \pi G T_{\mu \nu},
\end{eqnarray}
the above equation respects the energy momentum configuration
\begin{eqnarray}
T^t_t &=& T^{x_1}_{x_1} = T^{x_2}_{x_2} = 8\,  G \mu \, \delta^2(y_1,y_2)  e^{{4\over 3}E} \nonumber \\
T^{y_1}_{y_1}  &=& T^{y_2}_{y_2} = 0  \nonumber \\
T^{z_1}_{z_1} &=& ...= T^{z_6}_{z_6} = 0 \label{energy1}
\end{eqnarray}
The  energy momentum configuration (\ref{energy1})  is given by the brane action (\ref{brane1}) with the brane  conditions (\ref{branesplit1}-\ref{branesplit3}).  Analysing only the brane's  contribution of the energy momentum tensor we found  an essential aspect of this theory. Despite  the eleven dimensional space time, the energy momentum tensor is zero in a transverse plane.  This is a generalization of a  four dimensional  conical defect  for eleven dimensional supergravity.
We can consider  the redefinition as
 \begin{equation}
 \rho = { \Big({r \over r_0}\Big)^{1 - 4 G \mu} r_0 \over  (1 - 4  G \mu)}
\end{equation}
with the approximation $1- 8 G \mu \ln({r\over r_0})    \sim\Big({r\over r_0} \Big)^{-8G \mu } $ we have the metric bellow as
\begin{eqnarray}
ds^2 &=&f(\rho) \Big(  ds^2_{CD}+ dx_2^2  \Big) +g(\rho)ds_6^2 \label{scalar-tensor}
\end{eqnarray}
where for convenience we defined the metric with a dimensional conical defect sector given by  $ds_{CD}^2 $  and  the Euclidian sector  $ds_6^2$ is given by  \begin{eqnarray}
ds_{CD}^2 &=& - dt^2 + dx_1^2 +  d\rho^2 + \rho^2  d  \bar \theta^2 \\
ds_6^2 &=& dr'^2 + r'^2 d\Omega_5^2\label{cod6}
\end{eqnarray}
where $r' = \sqrt{z_1^2 + ...+dz_6^2}$
with $ f(\rho) = \Big[1 - 2 \Delta \ln{\rho \over \rho_0^*}\Big]^{-{2 \over 3}} $ and $ g(\rho) = \Big[1 - 2 \Delta \ln{\rho \over \rho_0^*}\Big]^{1/3} $ with  $\rho_0^* =  {r_0 / \kappa} $, $\kappa = (1 - 4G \mu) $ and   $ \Delta = { 4 G \mu \over 1- 4 G \mu} $ .
This metric corresponds to the conical defect with  warp factors.  
The  sector $ds_{CD}^2$ of this metric is locally known as Minkowiski with  $ \bar \theta = \kappa \theta$,  like a cosmic string in four dimensions.  If $ \kappa \leq 1$ or $\kappa \geq 1 $ we can write the range of the angular sector as $ 0 \leq \bar \theta  \leq 2 \pi \kappa$.   This conical geometry has an angular deficit given by $ \delta \theta = 8 \pi G \mu$ . 
The charge can be calculated using the same framework developed in \cite{Myers:1986un,Bousso:2000xa} and  by using the Dirac quantization we can prove that this charge is quantized.  
We also can see that the metric (\ref{scalar-tensor})  presents a warp factor \cite{Randall:1999ee}  similar to the ones in a scalar tensor theory in a weak field approximation  \cite{Guimaraes:1996ti}

\section{ The  $AdS_4 \times S^1 \times \mathbb{E}^6 $  near the horizon  limit}

In this section let us consider the near horizon limit AdS  for SCFT. We can write the metric (\ref{scalar-tensor}) as\begin{equation}
ds^2 = \Big({r \over r_0}\Big)^{4 \alpha} (-dt^2 + dx_1^2 + dx_2^2) + \Big({r \over r_0}\Big)^{-2 \alpha} d^2r  +  \Big({r \over r_0}\Big)^{-2 \alpha} (r^2d\theta^2 + ds_6^2)\label{metric4}
\end{equation}
where $\alpha  $ depends of brane tension $ \mu$ as $\alpha = {4 \over 3} G \mu $. We saw in last sections that the $r_0$ is related with the  minimum radial length.  The Euclidean metric $ ds_6^2$ is given by (\ref{cod6}).  Now let us relate this  radial length  with the low energy limit. The  decoupled limit is obtained by taking the 11 dimensional Planck length to zero $ l_p \rightarrow 0$, keeping the world volume energies fixed and taking the separation $ U^a \equiv r/l_p^b $  and   $ V^{a'} \equiv r'/l_p^{b'  }$   fixed.   With this limit we  put the parameter $\alpha $, $"a"$ and $ "b"$ as arbitrary and we analyzed here the conditions for AdS.  With this transformation the metric (\ref{metric4}) becomes
\begin{eqnarray}
ds^2 &=& {U^{4 \alpha a }  \over (l_p^{-b} \,  r_0)^{4 \alpha}}  (-dt^2 + dx_1^2 + dx_2^2)  +{a^2 U^{2a(1-\alpha) - 2} \over l_p^{-2b (1-\alpha) } r_0^{-2 \alpha}}  dU^2   +       { U ^{2 a(1- \alpha)} \over  l_p^{ -2b(1-\alpha)} r_0^{-2 \alpha}} \, d^2 \theta           \nonumber \\ 
&+&  {a'^2 U^{-2a\alpha } V^{2a\alpha}  \over l_p^{2b \alpha } r_0^{-2 \alpha}}  {V^{2(a' - 1) - 2 a \alpha} \over l_p^{-2b (1-\alpha) } }   dV^2 + { U ^{-2 a \alpha } V^{2a \alpha}\over  l_p^{ \,\, \, 2b\alpha } \, r_0^{-2 \alpha}}    { V ^{2 a' -2a\alpha   }  \over  l_p^{ -2b' } }   \, d^2 \Omega_6
\end{eqnarray}
Let us consider $ (l_p^{-b} \,  r_0)^{4 \alpha} =  {a^2r_0^{2 \alpha} \over l_p^{-2b(1-\alpha)}}  =R_{AdS}^2  $ then we get $a= 1/2 $.
It is easy to see the low energy $AdS_4$ near the horizon limit when $ \alpha  \rightarrow 1 $.  There is a non zero flux of the dual four-form field strength  on the   on $S^1$. In this limit  $a' = 1/2$,  and ${V/U \sim 1}$. The metric  is as follows
\begin{eqnarray}
ds^2& =& {U^{2 }  \over R_{AdS}^2}  (-dt^2 + dx_1^2 + dx_2^2)  +{ R_{AdS}^2 \over U^2}  dU^2 + (2 R_{AdS})^{2}  d^2 \theta \nonumber \\
&+& {dz^2} + R'^2d\Omega_5^2   \label{AdSmetric}
\end{eqnarray}
where $z= R'^2 \ln V$. We can see that this near horizon limit's metric  is  the product of an  Anti-de-Sitter space-time  in the form of $AdS_4 \times S^1 \times \mathbb{E}^6$. 
We can  analyze  the metric  (\ref{AdSmetric})  and  we can see that this limit corresponds to  a fix $ G\mu $  responsible for the AdS  near the horizon limit.  In this theory we have  a =1/2 and   b = 3/2, both compatible with the Maldacena analysis. The  AdS  radius  is given by $R_{\odot} = 2R_{AdS} =   l_p (2^5 \pi^2 N)^{1/6} = r_0 $ so we can  relate  the minimum radius of the circle with the scale of M2-brane.

\section{Discussions and Remarks }

In this work we analyzed the topological defect with a conical deficit  in eleven dimensions. In our prescription we found the BPS solution analyzing the  Killing spinor, gauge field and Einstein equations. We considered the space time with a (3,2,6) split where the co-dimension two represents the defect and  generates the flux.  We showed that the split is consistent with the M2-Brane configuration and that it  is possible to describe a topological defect energy density with the sigma models. This split gives us the BPS solutions for  Killing spinors without intersection\cite{Gauntlett:1997cv} or rotation branes \cite{Gauntlett:1998kc}. We analyzed the Einstein equations and showed that the configuration is  a general form for  conical defect  like a cosmic string with warp factors in eleven dimensions.  
Despite  similarities with the results of  theories with lower dimensions, it is  important to observe that  the  eleven dimensional supergravity theory  gives us a  different solution for  the  usual conical defect in gravitation. 
This solution presents the mass for length unit $ \mu$, that is the tension of the brane,  in the Planck scale. In a solution for this defect in ten dimensions, $G\mu \geq 10^{-3}  $, but in the cosmic string the quantity is $G\mu \leq 10^{-5}$ .
The other difference is that our solution presents a natural superconductivity  given by the presence of the BPS SUSY bound that admit fermionic superconductivity \cite{Davis:1997bs,Brax:2006yb}.
In our work we analyzed the possibility of the existence of this solution directly given by the M2-brane, but this was only the first step to understand our solution. Our prescription is general, consisting in a vast variety of applications .  Here we will discuss  some of these applications.  Nowadays there are many ways to get  low dimension prescriptions. Our framework  contains the AdS near  horizon  limit important for the holographic principle. We  analyzed this limit  and concluded   that the metric is  $AdS_4 \times S^1 \times  \mathbb{E}^6$.
The interesting feature  here is that this limit, considering the (3,\underline{2},6) split, is only  possible because of the existence of a topological defect.  This limit occurs to the value $G\mu = 7,5 \times 10^{-1}  $.  This value is compatible with the fact of the ten dimensions to be $ G\mu \geq 10^{-3} $.  
There are a few studies about intersecting and rotationg branes  \cite{Boonstra:1997dy} that obtain a configuration $AdS_3 \times  S^3 \times \mathbb{E}^5$.. In our framework we can obtain the $AdS_3$  near the horizon limit in the resulting theory after the compactification into ten dimensions.
We have a dimensional reduction of space-time and world volume in $D=11$ that reduces the combined type IIA supergravity-superstring field equations into $D=10$.  The split  (10,1) gives us $x^M=(x^{ \mu}, x^2)$,  where   $\mu = (0,1,3,...9)$.
We used the same definition standard \cite{Fund2} and consider that the dilaton  is related with the $ g_{22}$ component. After the reduction we got  $
g_{22} = e^{4 \phi / 3} \, \, \,  \mbox{with } \, \, \,  g_{MN} = e^{- \phi/6} g_{ \mu  \nu} $
and the gauge potential became,  $ A_{\mu \nu  2} = B_{\mu \nu } $. 
Considering the solution (\ref{metric4}), with our split we had after  the reduction  the dilaton and the 2-form of gauge given by $
B_{01} = \pm \, e^{-E} $ and   $  e^{\phi} = e^{{1\over 2}E} $
with the brane condition  $X^{\mu} = \xi^{\mu}$, $ \mu =0,1$, \, $ X^a = \mbox{constant}$ and $ Z^{\tilde m} = \mbox{constant}$.
We used the same procedure for  $AdS_4$ near the horizon limit of the last section and we found $ AdS_3 \times S_1 \times  \mathbb{E}^6$. We found the $AdS_3$ near the horizon  limit without intersection brane in ten dimensions. We can obtain  more complicated solutions by intersection branes \cite{Gauntlett:1997cv} or adding angular momentum \cite{Gauntlett:1998kc}, this is a subject to future works.   
The idea is to analyze the  possibility of intersecting branes with the defect appearing in the Minkowiski  sector  \cite{Bayona:2010sd}.  
Another  important  application is the study of strongly coupled condensed matter systems.  
The topological defects type cosmic strings can be viewed as vortex configuration  where the only difference is the coupling with gravitation. We had seen that the conical defect configuration in the $AdS_4$ can, via holography,  give us the $(2+1)$ models where the defect is preserved  on the boundary via metric\cite{BallonBayona:2013gx} . 
Therefore we believe that the vortices  in  eleven dimensions supergravity theory can give us an interesting prescription to the  study of the holographic principle mainly for graphene-like materials.  Another important method to understand these systems is the formulation developed by  \cite{O'Bannon:2007in,BallonBayona:2013gx}  with the study of conserved charges that are preserved on the boundary. In materials  graphene like has been shown that a vortex configuration  can give us a mass gap. But  if the material  is put into a cone shape there appear polarized currents  that resembles a gravitational deformation caused by the angular deficit like cosmic string \cite{Vozmediano:2010zz,Cortijo:2006xs,Cortijo:2006mh}. The idea now is to develop our framework to understand these effects among others.

\vspace{1 true cm} { \bf ACKNOWLEDGEMENT: } 
The author would like to thank the ICTP in Trieste (Italy). The work was partially supported by a CNPq/Brazil fellowship.

\end{document}